\documentclass[conference]{IEEEtran}
    \IEEEoverridecommandlockouts
    \usepackage{cite}
    \usepackage{amsmath,amssymb,amsfonts}
    \usepackage{algorithm}
    \usepackage{algpseudocode}
    \usepackage{graphicx}
    \usepackage{textcomp}
    \usepackage{xcolor}
    \usepackage{array}
    \usepackage{colortbl}
    \usepackage{multirow}
    \usepackage{subcaption}
    \def\BibTeX{{\rm B\kern-.05em{\sc i\kern-.025em b}\kern-.08em
        T\kern-.1667em\lower.7ex\hbox{E}\kern-.125emX}}

\begin{document}
    \title{SENTRY: A Streaming Edge Network Trust Engine for Real-Time Audio Deepfake Detection}

\author{\IEEEauthorblockN{Chirantan Banerjee, Soumita Chatterjee,
Bhargab B. Bhattacharya and Hafizur Rahaman}
\IEEEauthorblockA{Indian Institute of Engineering Science and Technology, Shibpur, India}}

    \maketitle

\begin{abstract}
Advances in real-time voice cloning and neural vocoders have enabled large-scale streaming deepfake and vishing attacks, while simultaneously exposing voice communication systems to denial-of-service threats. Existing Network Detection and Response (NDR) solutions rely on computationally intensive, GPU-based post-processing techniques that require buffered audio and exhibit limited robustness against unseen speech synthesis models. This paper presents \textbf{SENTRY}, a hardware-oriented, streaming trust engine for inline deepfake detection at the network edge. SENTRY employs an integer lifting-wavelet front end together with a bank of multiplier-free transforms, including wavelet-packet envelopes, Walsh--Hadamard sequency profiles, waveform micro-statistics, and long-term spectral features, designed so that acoustic analysis needs only integer add, shift, compare and bitwise operations. A supervised gradient-boosted tree classifier is combined with an unsupervised Mahalanobis out-of-distribution detector trained solely on authentic speech, and their outputs are fused through a lightweight logistic decision layer to produce graduated \textsc{log}, \textsc{challenge}, or \textsc{drop} actions. Under a closed-set protocol, SENTRY reaches \textbf{98.75\%} accuracy at a \textbf{1.25\%} equal-error rate (EER) on WaveFake, with perfect separation of HiFi-GAN and WaveGlow; its supervised track alone achieves an AUC of \textbf{0.9994} (fused system: 0.9993). On ASVspoof~2019~LA (attacks A01--A06) transmitted through an $8$\,kHz $\mu$-law telephony channel, it reaches \textbf{94.75\%} accuracy at a \textbf{5.05\%} EER, with a supervised-track AUC of \textbf{0.9948} (fused: 0.9884) and per-attack EERs below 7\%. In our prototype, the feature front end runs as bit-exact integer code on a single Arm Cortex-A53 core, and the classifier and fusion stage are synthesized for a Xilinx Zynq UltraScale+ MPSoC using no DSP slices, returning a decision 220\,ns after a feature vector is ready. These results indicate that deepfake screening can be placed inline without GPUs or floating-point hardware; generalization to unseen generators is left to future work.
\end{abstract}


\section{Introduction}

Voice phishing, or vishing, has become one of the most effective forms of
social engineering in modern telecommunications~\cite{triantafyllopoulos2025vishing}.
Unlike a suspicious email, a phone call carries tone, urgency and the simple
human habit of trusting a voice, which makes people far more likely to comply
before they stop to think~\cite{heiding2027evaluating}. Generative speech
models have removed the last practical barrier for attackers: a convincing
cloned voice costs almost nothing to produce and can be generated on demand. A
recent population-scale study reports that AI-generated voices reach
compliance rates above $36\%$ in emotionally charged scenarios, and that
several commercial voice systems are already cheap enough to make automated
vishing economically worthwhile~\cite{heiding2027evaluating}. The threat is no
longer hypothetical; it is active, inexpensive and spreading.

Compounding the problem, people cannot be relied upon to spot these calls
themselves. The same study found listeners correctly identified AI-generated
callers only about $70\%$ of the time, close to chance when judging text, and,
critically, participants who use AI tools frequently were no better at
detection than those who never had~\cite{heiding2027evaluating}. Independent
perceptual studies reach the same conclusion: humans are poor discriminators
of cloned speech, and training offers little
benefit~\cite{barrington2025poorly,mueller2022humanperception}. Familiarity
with generative AI confers no protection. This makes automated, in-network
detection not merely convenient but necessary, because the human in the loop
is not a dependable line of defence.

The danger is not limited to fooling a single victim. When synthetic voices
are generated in bulk and pointed at corporate customer-service lines, they
can exhaust human agents and automated menus alike, tipping a call centre into
an operational denial-of-service against real customers. Defending against
this needs more than accuracy: it needs a decision made \emph{during} the
call. A vishing call is a live audio stream, not a file waiting on a disk, and
by the time a recording could be pulled aside and analysed offline, the
conversation is already over and the money already gone. A useful defence has
to render a verdict in the first seconds of a call, close to where the audio
enters the network.

This is where current detectors fall short, in three ways. First, the
strongest systems, such as AASIST~\cite{jung2022aasist} and
RawNet2~\cite{tak2021rawnet2}, and the self-supervised front ends that now
dominate the leaderboards~\cite{tak2022wav2vec,baevski2020wav2vec}, are
floating-point neural networks that rely on dense multiply-accumulate
hardware such as GPUs, an assumption that does not hold for a low-cost inline
device at a network gateway. Second, these models are trained against a fixed
list of known synthesisers and generalise poorly when a new one appears.
M\"uller et~al.~\cite{mueller2022generalize} showed that detectors tuned on
ASVspoof degrade by up to an order of magnitude on found, real-world audio,
which suggests the field has fitted itself rather closely to one
benchmark~\cite{yi2023survey}. Third, existing tools are built as offline
forensic utilities that look only at the waveform, ignoring the live
network-layer signals, such as RTP jitter and packet-arrival volatility, that
a real-time system sitting in the data path could exploit. In short, deepfake
defence today is an after-the-fact analysis, not the real-time inline network
security function an NDR appliance actually needs. SENTRY addresses the third
gap through inline deployment but currently uses acoustic features only;
exploiting network-layer signals is left to future work.

\begin{figure}[t]
        \centering
        \includegraphics[width=0.89\linewidth]{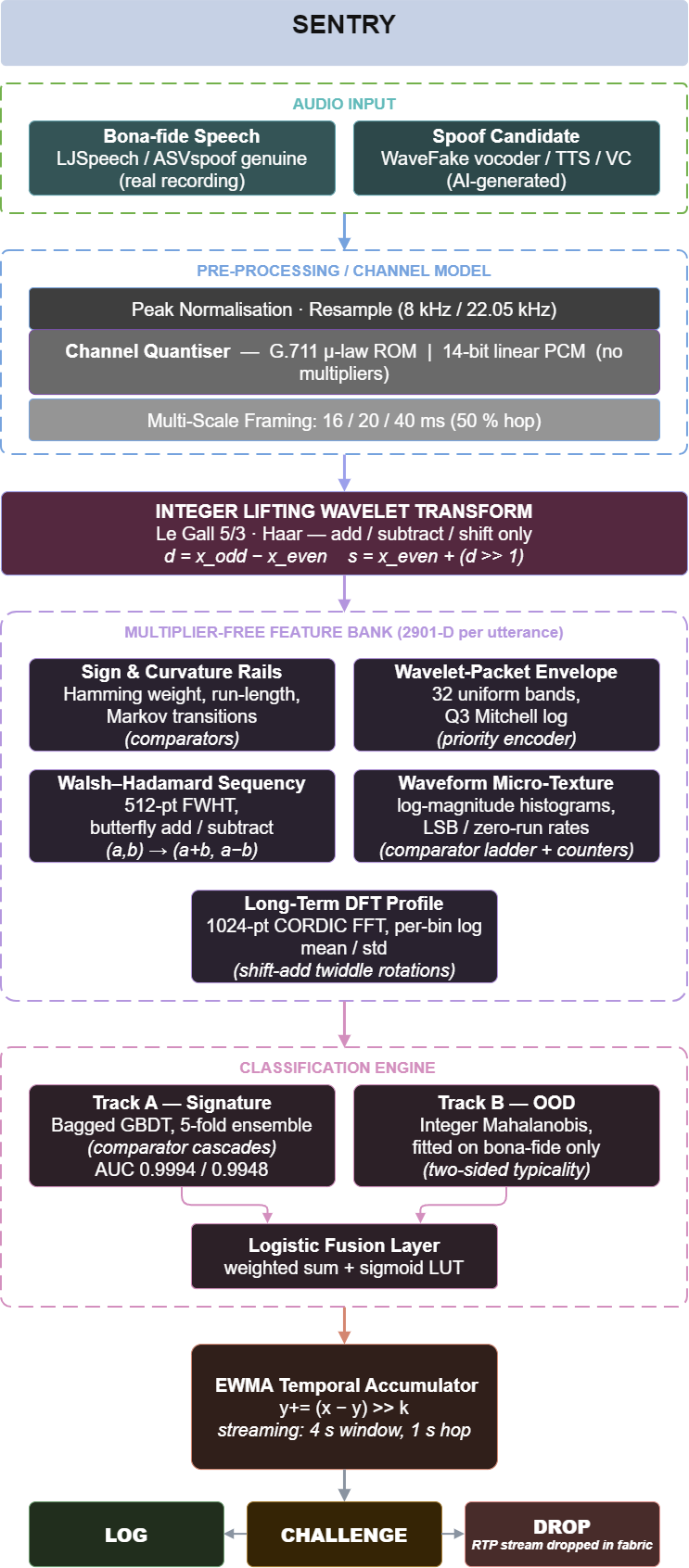}
    \caption{Overview of the proposed framework.}
    \label{block_dig}
    \vspace{-10pt}
\end{figure}

We present \textbf{SENTRY} (Streaming Edge-Network Trust Engine;
Fig.~\ref{block_dig}), a lightweight, hardware-oriented detector built to
close these gaps. The central idea is to do all detection work with integer
additions, shifts, comparisons, \textsc{xor} and popcount operations, so that
the pipeline can be mapped onto FPGA look-up tables with no floating-point
units and no DSP multipliers. On that arithmetic foundation, SENTRY makes
three design choices aimed directly at the three failures above:
\begin{itemize}
    \item \textbf{Artefact-oriented features.} Rather than a mel front end,
    which compresses the very band where vocoder artefacts live, SENTRY uses a
    multi-resolution integer wavelet decomposition together with a fine
    long-term spectral profile, so the high-frequency evidence survives into
    the classifier.
    \item \textbf{Built for live audio.} SENTRY is designed to consume an RTP
    stream directly and produce a running human-versus-AI verdict within the
    first seconds of a call, rather than waiting for a completed recording.
    \item \textbf{Deployable at the edge.} Every operation on the inference
    path is designed as a cheap integer primitive, so the detector can be
    mapped to FPGA fabric and run inline without a GPU in the loop.
\end{itemize}

We prototype SENTRY on a Xilinx Zynq UltraScale+ MPSoC. In this prototype,
the feature front end runs as bit-exact integer C on one Arm Cortex-A53 core
of the processing system, and the classifier and fusion stage run as a Vitis
HLS kernel in the programmable logic. Feature vectors are transferred to the
fabric via AXI-DMA, the \textsc{log}/\textsc{challenge}/\textsc{drop} verdict
is computed without further CPU involvement, and the design allows the
offending stream to be dropped directly in hardware. Trained parameters load
at start-up over AXI-Lite and can be refreshed without rebuilding the
bitstream.

\section{Related Work}
\label{sec:related}

\subsection{Neural Countermeasures and Their Cost}
The anti-spoofing literature has converged on end-to-end neural models. RawNet2
consumes raw waveforms through learnable sinc filters~\cite{tak2021rawnet2},
AASIST models spectral and temporal artefacts jointly with graph
attention~\cite{jung2022aasist}, and the current strongest systems fine-tune
large self-supervised encoders such as wav2vec~2.0 on spoofing
data~\cite{tak2022wav2vec,baevski2020wav2vec}, often with heavy augmentation
to survive channel variation~\cite{tak2022rawboost}. These deliver excellent
in-domain numbers on ASVspoof~2019~\cite{wang2020asvspoof2019} and
2021~\cite{yamagishi2021asvspoof2021}. Their sizes range from hundreds of
thousands of parameters (AASIST) to hundreds of millions (self-supervised
encoders), but all of them depend on dense floating-point multiply-accumulate
hardware. Nothing in that design admits a multiplier-free, line-rate
implementation, which is the constraint SENTRY takes as primary.

\subsection{The Generalisation Problem}
A second and more troubling thread is that these numbers do not travel.
M\"uller et~al.~\cite{mueller2022generalize} evaluated a suite of published
detectors on found, in-the-wild recordings of public figures and observed
error-rate increases of up to an order of magnitude relative to ASVspoof,
concluding that the field may have tailored its solutions too closely to a
single benchmark. Surveys of the area report the same
pattern~\cite{yi2023survey}, and successive challenge editions have added
codec, compression and channel variability precisely because in-domain
accuracy had ceased to be informative~\cite{yamagishi2021asvspoof2021}. This
is why we report per-generator behaviour and two acoustically different
channel conditions rather than a single pooled figure.

\subsection{Lightweight and Applied Voice-Scam Detectors}
Closest to our deployment goal are recent applied systems. Saxena
et~al.~\cite{saxena2025aipowered} pair a CNN with a recurrent network over
MFCC and spectrogram inputs and add SHAP/LIME explainability, reporting $98\%$
accuracy for real-versus-fake audio. Rubika et~al.~\cite{rubika2026voice}
combine a 1D-CNN branch with a Speech-Transformer branch to capture prosodic
urgency and conversational intent, reporting a $94.8\%$ F1-score and
explicitly targeting lightweight deployment. Both frame the problem correctly,
as a live telecommunications threat rather than an offline forensic exercise.

Three gaps remain, and SENTRY is designed around them. \emph{Scale and
evaluation:} these systems are trained and tested on small, self-collected
corpora ($1{,}000$ and $1{,}200$ clips respectively), with a test partition of
a few tens of clips in the first case; neither performs cross-dataset
validation, and Rubika et~al. name the small dataset and the absence of
cross-corpus evaluation as principal limitations of their own study. Given the
degradation documented in~\cite{mueller2022generalize}, accuracy measured on
such partitions is hard to transfer to an operational setting.
\emph{Front-end choice:} both rely on mel-scaled features, which compress
precisely the high-frequency band where neural-vocoder upsampling artefacts
accumulate; constant-Q and cepstral front ends share this
tendency~\cite{todisco2017cqcc}. \emph{Deployment reality:} the CNN--RNN
system is served from a Flask/React stack with Twilio call capture and cloud
hosting, so detection still runs on general-purpose servers, while the
Transformer--CNN hybrid is designed for edge feasibility but defers parameter
count, memory footprint and latency measurement to future work. In both cases
the arithmetic is floating-point neural computation, which needs DSP blocks or
a GPU.

SENTRY differs on all three axes. It is evaluated on two established public
corpora spanning distinct generator families---neural vocoders and classical
TTS/voice-conversion attacks---rather than a single self-collected set, with
per-generator breakdowns reported for every attack system. It replaces the mel
front end with a multi-resolution integer lifting wavelet plus a fine
long-term spectral profile that preserves the high-frequency evidence. And it
executes entirely in integer arithmetic, so the classifier is not merely
``lightweight'' in parameter count but structurally free of multipliers and
synthesisable into FPGA fabric inline with the RTP path.

\section{Methodology}

SENTRY is a streaming trust engine that decides whether a telephony or VoIP
audio stream carries genuine human speech or machine-generated audio, using
integer additions, subtractions, shifts, comparisons, \textsc{xor} and
popcount operations, so that the inference datapath is designed to map to
FPGA lookup logic with no DSP multipliers. An incoming stream first passes
through a channel front end (an $8$\,kHz $\mu$-law telephony model, or a
wideband 14-bit linear quantizer for Opus/G.722-grade links). The integer PCM
is then analysed by a bank of multiplier-free transforms at several time
scales, each turning the waveform into a compact set of statistics that expose
a different kind of synthesis artefact. These statistics feed two parallel
decision tracks: Track~A, a gradient-boosted tree
ensemble~\cite{ke2017lightgbm} trained on known bona-fide and spoof examples,
and Track~B, an integer Mahalanobis out-of-distribution
detector~\cite{lee2018mahalanobis} fitted on genuine speech only. A
lightweight logistic layer fuses the two track scores, an
exponentially-weighted accumulator smooths the decision over time, and a
graduated policy emits one of three actions---\textsc{log}, \textsc{challenge}
or \textsc{drop}---at operator-chosen false-alarm rates. Training happens
offline in floating point; only integer thresholds and shift-quantized
matrices are deployed.

The feature front end is a stack of complementary multiplier-free transforms;
each was added to capture an artefact class the previous ones could not see.
We describe them in the order signal information flows.

\subsection{Integer Lifting Wavelet and Sign Rails}
Each frame is decomposed with a multi-level integer lifting wavelet transform
(Le~Gall 5/3~\cite{legall1988subband} or Haar), realised through the lifting
factorisation~\cite{daubechies1998lifting}, which maps integers to integers
using only add, subtract and arithmetic shift~\cite{calderbank1998integer}.
For the Haar case the lifting step is
\begin{equation}
d = x_{\mathrm{odd}} - x_{\mathrm{even}}, \qquad
s = x_{\mathrm{even}} + \lfloor d / 2 \rfloor ,
\end{equation}
where $s$ and $d$ are the approximation and detail sub-bands and
$\lfloor\cdot/2\rfloor$ is a right shift. From the wavelet coefficients we
form binary ``sign-of-derivative'' rails: bit $b_t=1$ if a coefficient
increased relative to the previous frame, and a second curvature rail marking
sign changes of the second temporal difference. These rails are
magnitude-blind by construction, cost only comparators, and summarise the fine
temporal dynamics of the coefficients through per-band Hamming weights,
run-length statistics and first-order Markov transition rates.

\subsection{Wavelet-Packet Envelope}
The sign rails cannot see spectral \emph{magnitude} structure, so we add a
wavelet-packet filterbank: the top octave is split further with repeated Haar
lifting to yield $32$ uniform sub-bands, an integer, multiplier-free analogue
of a linear-frequency filterbank. For each band we take a fixed-point log
energy using Mitchell's approximation~\cite{mitchell1962logarithm},
\begin{equation}
\log_2 v \approx e + \mathrm{frac}, \qquad e=\lfloor \log_2 v \rfloor ,
\end{equation}
where $e$ is the leading-one position (a priority encoder) and the fractional
bits are a barrel shift---no multiplier. Band statistics are computed over all
frames and, separately, over high-energy (voiced) frames selected by an energy
comparator, since silence dilutes the envelope.

\subsection{Walsh--Hadamard Sequency Profile}
To resolve narrow comb structure that a shallow packet tree smears, we apply a
Fast Walsh--Hadamard Transform (FWHT), the classic multiplier-free spectral
transform~\cite{beauchamp1984walsh}, built entirely from butterfly
add/subtract stages:
\begin{equation}
(a,b)\;\longmapsto\;(a+b,\;a-b),
\end{equation}
recursively over $\log_2 N$ stages. Per-bin means and standard deviations of
the integer log-magnitude sequency spectrum, taken over voiced frames, give a
fine texture descriptor for the mid band. Averaged over voiced frames,
synthetic speech sits consistently below genuine speech across most sequency
bands, and it is this systematic offset rather than any single bin that the
downstream classifier exploits.

\subsection{Waveform Micro-statistics}
Real recordings carry a microphone/pre-amp noise floor that dithers the
low-order sample bits; vocoder output does not. We therefore histogram the
log-magnitudes of the samples and of their first differences and record
least-significant-bit and zero-run rates---a comparator ladder plus a handful
of counters in hardware---to expose this ``too clean'' fine-amplitude texture.

\subsection{Long-Term DFT Log-Magnitude Profile}
The most discriminative descriptor is a fine Discrete Fourier Transform of
long frames. The average energy per frequency bin is precisely where the
original WaveFake analysis locates the real-versus-fake
difference~\cite{frank2021wavefake}, and it is concentrated in the higher
frequencies. For a frame $x$ we compute
\begin{equation}
X(k)=\sum_{n=0}^{N-1} x(n)\,w(n)\,e^{-j 2\pi kn/N},
\end{equation}
and store per-bin means and standard deviations of the integer log magnitude
over voiced frames. In hardware this is the golden model of a CORDIC FFT,
whose twiddle rotations are shift-add iterations, so the multiplier-free
property is preserved; a quantization ablation of the fixed-point path is
left to future work.

\subsection{Two-Track Decision and Fusion}
Track~A is a bagged $k$-fold gradient-boosted tree
ensemble~\cite{ke2017lightgbm} over the full feature vector. Track~B is an
integer Mahalanobis score fitted on bona-fide speech only, using robust
median/MAD scaling, a shrinkage covariance estimator and a two-sided
log-distance so that ``suspiciously typical'' (over-smoothed) audio is flagged
as strongly as ``far'' audio:
\begin{equation}
D(f)=(f-\mu)^{\top}\,\Sigma^{-1}\,(f-\mu),\qquad
z=\frac{|\log(1+D)-m|}{\mathrm{MAD}} .
\end{equation}
A two-input logistic layer fuses the track scores; in RTL this is a weighted
sum plus a small sigmoid lookup. Algorithm~\ref{alg:sentry} summarises the
inference path.

\begin{algorithm}[t]
\caption{SENTRY inference for one stream}
\label{alg:sentry}
\begin{algorithmic}[1]
\State $p \gets \textsc{ChannelQuantize}(x)$
\State $F \gets [\,]$
\For{each analysis scale}
  \State compute lifting-wavelet sign/curvature rail statistics
  \State compute $32$-band packet log-energy envelope
  \State compute FWHT sequency profile and waveform micro-statistics
  \State compute long-term DFT log-magnitude profile
  \State append all to $F$
\EndFor
\State $s_A \gets \textsc{TrackA}(F)$;\quad $s_B \gets \textsc{TrackB}(F)$
\State $s \gets \textsc{Fuse}(s_A,s_B)$;\quad
       $\bar{s} \gets \textsc{Ewma}(s)$
\State \Return $\textsc{Policy}(\bar{s})\in\{\textsc{log},
       \textsc{challenge},\textsc{drop}\}$
\end{algorithmic}
\end{algorithm}

The novelty of SENTRY is that it reaches high closed-set accuracy with a
feature front end designed around integer primitives: every transform in the
inference path, from the lifting wavelet to the CORDIC-FFT profile, is built
from adds, shifts and comparators, so that it maps to LUT logic without DSP
slices. Two kinds of statistic still involve products of signal-dependent
values, namely the per-bin standard deviations and the Mahalanobis quadratic
form; in this work they are computed in the integer golden model, and their
multiplier-free hardware realisation (for example, small squaring look-up
tables or log-domain squaring) is part of the front-end synthesis left to
future work. The feature set is also deliberately artefact-oriented rather
than model-specific: the packet and DFT profiles describe \emph{how}
synthesised speech deviates from natural speech---high-frequency energy
imbalance, over-smoothed spectral envelopes, an absent noise floor---which are
properties shared across broad generator families rather than tied to one
architecture. Track~B, fitted on genuine speech alone, is included as a
mechanism for flagging inputs that leave the bona-fide manifold;
Section~\ref{sec:results} reports its measured contribution, including where
it underperforms.

The datapath is designed to run in real time on a mid-range FPGA at the
network edge, inline with a VoIP gateway. Because the transforms avoid
multipliers, the front end maps to a pipeline of the fabric's native
primitives: the $\mu$-law stage is a $256$-entry ROM, the lifting wavelet and
Haar packet tree are add/subtract/shift stages, the FWHT is a butterfly
network, the DFT profile is a CORDIC FFT, and the per-band log is a priority
encoder plus barrel shifter. The tree ensemble becomes comparator cascades and
the fusion layer a short weighted sum with a sigmoid lookup, so the classifier
fits in on-chip block RAM and LUTs (Section~\ref{subsec:res_hw}) and
comfortably meets the $20$--$40$\,ms frame budget.

\section{Results and Discussions}
\label{sec:results}

We evaluate SENTRY on two established public corpora chosen to span different
generator families. WaveFake~\cite{frank2021wavefake} supplies six neural
vocoders (five GAN-based and one flow-based) resynthesising LJSpeech, and is
analysed at $22.05$\,kHz through the wideband integer front end, which retains
the $8$--$11$\,kHz band where vocoder artefacts are strongest.
ASVspoof~2019~LA~\cite{wang2020asvspoof2019} supplies six text-to-speech and
voice-conversion attack systems over many speakers, and is analysed at
$8$\,kHz through the G.711 $\mu$-law telephony channel model, matching the
narrowband PSTN condition an inline appliance would actually see.
Each dataset contributes $8{,}000$ utterances, divided into disjoint
training and test sets in which every attack system appears in both; for
ASVspoof~2019~LA they are drawn from the training and development partitions
(attacks A01--A06). This closed-set protocol differs from the official
ASVspoof~2019~LA evaluation partition, whose attacks A07--A19 are unseen in
training; we leave that protocol, and a same-split comparison with published
neural countermeasures, to future work. All feature extraction and inference
use integer arithmetic only; the floating-point transforms serve purely as
golden-model references. We report equal-error rate alongside accuracy, F1
and AUC, following standard practice in the anti-spoofing literature;
confidence intervals are not yet reported, so small differences should be
read with caution.

\subsection{Detection Performance Across Datasets}
\label{subsec:res_main}

Table~\ref{tab:main_results} reports both datasets with each decision track
shown separately, and the best value in every column is set in bold. On
WaveFake the fused system reaches $98.75\%$ accuracy at a $1.25\%$ equal-error
rate. On ASVspoof~2019~LA the same pipeline, operating through a narrowband
telephony channel that discards everything above $4$\,kHz, reaches $94.75\%$
accuracy at a $5.05\%$ equal-error rate. We attribute much of the gap between
the two to the channel: the wideband condition preserves the high-frequency
band the packet and DFT profiles depend on, while the $\mu$-law trunk removes
it. The two datasets also differ in speakers and generator families, so this
comparison is indicative rather than controlled. The ASVspoof figure should be
read as the harder, more operationally realistic PSTN operating point, since
channel and codec variability is exactly what later challenge editions were
designed to stress~\cite{yamagishi2021asvspoof2021}.

The per-track breakdown deserves a plain statement. Track~A carries the
discriminative load on both datasets and achieves the
\emph{best ranking quality of any configuration}, with an AUC of $0.9994$ on
WaveFake and $0.9948$ on ASVspoof, higher in both cases than the fused score.
The fusion layer trades a little ranking quality for a slightly better
decision threshold: fusion lowers EER from $1.31\%$ to $1.25\%$ on WaveFake
and from $5.38\%$ to $5.05\%$ on ASVspoof. These differences are small, and
without confidence intervals we do not claim they are significant. Track~B,
the unsupervised out-of-distribution engine, performs weakly in this
closed-set regime ($29.89\%$ and $43.18\%$ EER), which is unsurprising: with
every generator represented in training, a supervised model leaves little for
an anomaly detector to add, and a single-Gaussian model of the bona-fide
manifold is a coarse description of a high-dimensional feature space. We
retain Track~B because its intended value lies in the unseen-generator regime
that a closed-set protocol does not exercise~\cite{mueller2022generalize};
quantifying that value under leave-one-generator-out and cross-corpus
protocols is ongoing work.

\begin{table}[t]
\centering
\caption{Closed-Set Detection Performance on Both Datasets
($8{,}000$ utterances each, integer datapath). Best value per column in bold.}
\label{tab:main_results}
\setlength{\arrayrulewidth}{0.4pt}
\renewcommand{\arraystretch}{1.25}
\setlength{\tabcolsep}{6pt}
\footnotesize
\begin{tabular}{|p{2.4cm}|l|c|c|c|c|}
\hline
\rowcolor[gray]{0.82}
\textbf{Dataset} & \textbf{Track} &
\shortstack[c]{\textbf{Acc.}\\\textbf{(\%)}} &
\shortstack[c]{\textbf{F1}\\\textbf{(\%)}} &
\textbf{AUC} &
\shortstack[c]{\textbf{EER}\\\textbf{(\%)}} \\
\hline
WaveFake
& Fused   & \textbf{98.75} & \textbf{98.75} & 0.9993 & \textbf{1.25} \\
\cline{2-6}
\rowcolor[gray]{0.95}
& Track A & 98.69 & 98.69 & \textbf{0.9994} & 1.31 \\
\cline{2-6}
& Track B & 70.11 & 70.16 & 0.7195 & 29.89 \\
\hline
ASVspoof 2019 LA (A01--A06)
& Fused   & 94.75 & 97.00 & 0.9884 & 5.05 \\
\cline{2-6}
\rowcolor[gray]{0.95}
& Track A & 94.17 & 96.65 & \textbf{0.9948} & 5.38 \\
\cline{2-6}
& Track B & 56.83 & 70.23 & 0.6024 & 43.18 \\
\hline
\end{tabular}
\vspace{-0.2cm}
\end{table}

\subsection{Per-Generator Detection}
\label{subsec:res_per_model}

Table~\ref{tab:per_model} breaks the fused score down by the AI model that
produced the audio. This is the view an operator cares about, since it answers
which synthesisers the detector actually catches. Two generators, HiFi-GAN and
WaveGlow, are separated without error at the equal-error operating point, and
Parallel WaveGAN and MelGAN are close behind. Multi-band MelGAN, which
synthesises sub-bands separately, and Full-band MelGAN are the hardest of the
GAN family. On the ASVspoof side the per-attack EERs of the six attack systems
range from $1.60\%$ to $6.95\%$ despite the narrowband channel, with A06, a
voice-conversion system, the most difficult. Per-generator accuracy is
measured at each generator's own EER threshold (Table~\ref{tab:per_model},
note), so it shows separability rather than the detection rate at a single
deployed threshold.

\begin{table}[t]
\centering
\caption{Detection Accuracy by AI Model (fused score, each generator
versus bona fide)}
\label{tab:per_model}
\setlength{\arrayrulewidth}{0.4pt}
\renewcommand{\arraystretch}{1.25}
\setlength{\tabcolsep}{3pt}
\footnotesize
\begin{tabular}{|l|c|c|c|c|}
\hline
\rowcolor[gray]{0.82}
\textbf{AI Model} & \textbf{Type} &
\shortstack[c]{\textbf{Accuracy of} \\ \textbf{Being AI (\%)}$^{a}$} &
\textbf{AUC} &
\shortstack[c]{\textbf{EER} \\ \textbf{(\%)}} \\
\hline
\multicolumn{5}{|l|}{\emph{WaveFake ($22.05$\,kHz wideband)}} \\
\hline
HiFi-GAN            & GAN  & \textbf{100.00} & \textbf{1.0000} & \textbf{0.00} \\
\rowcolor[gray]{0.95}
WaveGlow            & Flow & \textbf{100.00} & \textbf{1.0000} & \textbf{0.00} \\
Parallel WaveGAN    & GAN  & 99.87 & 1.0000$^{b}$ & 0.13 \\
\rowcolor[gray]{0.95}
MelGAN              & GAN  & 99.36 & 1.0000$^{b}$ & 0.64 \\
Multi-band MelGAN   & GAN  & 98.43 & 0.9981 & 1.57 \\
\rowcolor[gray]{0.95}
Full-band MelGAN    & GAN  & 98.07 & 0.9980 & 1.93 \\
\hline
\multicolumn{5}{|l|}{\emph{ASVspoof 2019 LA ($8$\,kHz G.711 telephony)}} \\
\hline
A02                 & TTS & 98.40 & 0.9902 & 1.60 \\
\rowcolor[gray]{0.95}
A01                 & TTS & 98.11 & 0.9966 & 1.89 \\
A05                 & VC  & 97.31 & 0.9912 & 2.69 \\
\rowcolor[gray]{0.95}
A04                 & TTS & 95.59 & 0.9873 & 4.41 \\
A03                 & TTS & 95.42 & 0.9975 & 4.58 \\
\rowcolor[gray]{0.95}
A06                 & VC  & 93.05 & 0.9686 & 6.95 \\
\hline
\end{tabular}
\\[2pt]
\raggedright\scriptsize
$^{a}$Measured at each generator's own EER threshold, so it equals
$100-\mathrm{EER}$.
$^{b}$Rounded to four decimals; the unrounded value is below 1, consistent
with $\mathrm{AUC}\le 1-\mathrm{EER}^2$.
\vspace{-0.3cm}
\end{table}

\subsection{FPGA Implementation and Real-Time Behaviour}
\label{subsec:res_hw}

The classifier and fusion stage were described in Vitis HLS and synthesised
for the Zynq UltraScale+ target; in this prototype the feature front end runs
as bit-exact integer C on the processing system. The front end is designed for
the same integer primitives: the channel model is a small ROM, the lifting
wavelet and packet tree are add/subtract/shift
stages~\cite{daubechies1998lifting}, the Walsh--Hadamard stage is a butterfly
network~\cite{beauchamp1984walsh}, the long-term profile is a CORDIC-style
shift-add block, and the per-band log is a priority encoder with a barrel
shifter~\cite{mitchell1962logarithm}; synthesising it is future work. On a
positive verdict, the offending RTP stream can be dropped in the fabric with no
host CPU involvement, which is the inline mitigation an NDR appliance
requires. This is the concrete difference from cloud-served detectors, where
audio must be captured through a telephony API, shipped to a server and
classified there before any action can be taken~\cite{saxena2025aipowered},
and from edge-oriented neural hybrids whose latency and memory cost on
embedded hardware remain unmeasured~\cite{rubika2026voice}.

Table~\ref{tab:resource} reports post-synthesis resource utilisation of the
classifier and fusion kernel: it consumes 11,656 LUTs (5.06\% of the 230,400
available), 1,090 LUTRAMs (1.07\%), 9 BRAMs (2.88\% of 312), and 13,249
registers (2.88\% of 460,800), while requiring zero DSP slices, confirming
that this kernel is realised entirely from integer compare and add logic
rather than multiply-accumulate resources. Table~\ref{tab:timing} reports its
timing and power: at a 100~MHz clock the kernel returns a decision 220~ns
(22 cycles) after a feature vector is ready, corresponding to 4.55 million
feature vectors per second. The 2.94~W figure is the Vivado on-chip power
estimate for the whole device, including the processing system and static
power, rather than the incremental power of the kernel. End-to-end time to
first verdict is dominated by the window of audio the policy accumulates
rather than by this compute latency, and is not measured here.

\begin{table}[t]
\centering
\caption{Post-Synthesis Resource Utilization of the Classifier and Fusion
Kernel (Zynq UltraScale+ XCZU7EV)}
\label{tab:resource}
\setlength{\arrayrulewidth}{0.4pt}
\renewcommand{\arraystretch}{1.25}
\setlength{\tabcolsep}{3pt}
\small
\begin{tabular}{|l|c|c|c|}
\hline
\rowcolor[gray]{0.82}
\textbf{Resource} & \textbf{Used} & \textbf{Available} & \textbf{Utilization\,(\%)} \\
\hline
LUT               & 11,656 & 230,400 & 5.06 \\
\hline
\rowcolor[gray]{0.95}
LUTRAM            & 1,090  & 101,760 & 1.07 \\
\hline
BRAM              & 9      & 312     & 2.88 \\
\hline
\rowcolor[gray]{0.95}
Register          & 13,249 & 460,800 & 2.88 \\
\hline
DSP               & \textbf{0} & 1,728 & \textbf{0.00} \\
\hline
\end{tabular}
\vspace{-0.3cm}
\end{table}

\begin{table}[t]
\centering
\caption{Timing and Power of the Classifier and Fusion Kernel (power is the
Vivado on-chip estimate for the whole device, including the processing
system)}
\label{tab:timing}
\setlength{\arrayrulewidth}{0.4pt}
\renewcommand{\arraystretch}{1.25}
\setlength{\tabcolsep}{3pt}
\small
\begin{tabular}{|c|c|c|c|}
\hline
\rowcolor[gray]{0.82}
\textbf{Clock (MHz)} & \textbf{Latency (ns)} &
\shortstack[c]{\textbf{Throughput}\\\textbf{(M vectors/s)}} & \textbf{Power (W)} \\
\hline
100 & 220 & 4.55 & 2.94 \\
\hline
\end{tabular}
\end{table}

\subsection{Comparison with Prior Systems}

Table~\ref{tab:compare_prior} places SENTRY beside representative prior
detectors. We compare design points rather than headline accuracies, because
the reported figures come from different corpora and, in the case of the
CNN--RNN and Transformer--CNN systems, from small self-collected test
partitions that are not commensurable with the public benchmarks used
here~\cite{saxena2025aipowered,rubika2026voice}. What the table makes visible
is the combination SENTRY targets: a front end that keeps high-frequency
evidence, arithmetic that needs no multipliers, evaluation on more than one
public corpus with per-generator reporting, and a deployment target that sits
in the packet path rather than behind it.

\begin{table}[t]
\centering
\caption{Comparison of Design Points with Representative Prior Systems}
\label{tab:compare_prior}
\setlength{\arrayrulewidth}{0.4pt}
\renewcommand{\arraystretch}{1.25}
\setlength{\tabcolsep}{2pt}
\scriptsize
\begin{tabular}{|l|c|c|c|c|}
\hline
\rowcolor[gray]{0.82}
\textbf{Method} &
\textbf{Front End} &
\textbf{Arithmetic} &
\shortstack[c]{\textbf{Public multi-} \\ \textbf{corpus eval}} &
\shortstack[c]{\textbf{Deploy} \\ \textbf{target}} \\
\hline
CNN--RNN + XAI~\cite{saxena2025aipowered} & MFCC/Spec & Float NN & No & Cloud \\
\hline
\rowcolor[gray]{0.95}
Transformer--CNN~\cite{rubika2026voice}   & Mel-spec  & Float NN & No & Mobile$^{*}$ \\
\hline
RawNet2~\cite{tak2021rawnet2}             & Raw/sinc  & Float NN & Yes & GPU \\
\hline
\rowcolor[gray]{0.95}
AASIST~\cite{jung2022aasist}              & Raw/graph & Float NN & Yes & GPU \\
\hline
wav2vec2-AASIST~\cite{tak2022wav2vec}     & SSL       & Float NN & Yes & GPU \\
\hline
\rowcolor[gray]{0.95}
\textbf{SENTRY (ours)} & \shortstack[c]{Integer\\lifting DWT} &
\shortstack[c]{Integer\\(no DSP)} & \textbf{Yes} &
\shortstack[c]{FPGA\\Online} \\
\hline
\end{tabular}
\\[2pt]
\raggedright\scriptsize $^{*}$Stated design goal; latency and memory on
embedded hardware are left to future work in the cited study.
\vspace{-0.4cm}
\end{table}

\section{Conclusion}

This paper presented \textbf{SENTRY}, a streaming deepfake voice detector built entirely on integer arithmetic. By combining an integer lifting-wavelet front end, multiplier-free feature transforms, and a dual-track decision engine, SENTRY is designed for inline deployment on FPGA-based network infrastructure without floating-point units or DSP multipliers; in our prototype the classifier and fusion stage are synthesised with zero DSP slices, while the front end runs as integer code on the processing system. Under a closed-set protocol, SENTRY achieves \textbf{98.75\%} accuracy with a \textbf{1.25\%} equal-error rate (EER) on WaveFake, perfectly separating HiFi-GAN and WaveGlow, and \textbf{94.75\%} accuracy with a \textbf{5.05\%} EER on ASVspoof~2019~LA (attacks A01--A06) transmitted over an $8$\,kHz $\mu$-law telephony channel, with per-attack EERs below 7\%. The supervised detection path delivers the best ranking performance, with AUCs of \textbf{0.9994} and \textbf{0.9948}, while the fusion stage slightly improves operating-point selection. We attribute the narrowband performance reduction primarily to the loss of high-frequency spectral information. The out-of-distribution track is weak in the closed-set regime, and generalisation to unseen generators, which motivates it, remains to be demonstrated. Future work will evaluate SENTRY on the official ASVspoof~2019~LA evaluation partition and under leave-one-generator-out and cross-corpus protocols with same-split neural baselines, synthesise the complete front end, assess robustness to codecs and adaptive attacks, and explore network-layer evidence such as RTP timing.

\vspace{-0.2cm}

{\footnotesize
\bibliography{arxiv_ref}
\bibliographystyle{IEEEtran.bst}
}
\end{document}